\documentclass[prb, aps, twocolumn, superscriptaddress,noshowpacs]{revtex4}
\usepackage[english]{babel} 
\usepackage[english]{layout}
\usepackage{amsmath,amsfonts,amssymb}
\usepackage{graphicx}
\usepackage{mathrsfs}
\usepackage{float}
\usepackage{color}
\usepackage{braket}
\usepackage{version}
\usepackage{array,multirow,makecell}
\usepackage[T1]{fontenc}

\begin{document}

\title{Soliton formation in an exciton-polariton condensate at a bound state in the continuum}
\author{I. Septembre}
\affiliation{Universit\'e Clermont Auvergne, Clermont Auvergne INP, CNRS, Institut Pascal, F-63000 Clermont-Ferrand, France}
\author{I. Foudjo}
\affiliation{Universit\'e Clermont Auvergne, Clermont Auvergne INP, CNRS, Institut Pascal, F-63000 Clermont-Ferrand, France}
\author{V. Develay}
\affiliation{Laboratoire Charles Coulomb, Université de Montpellier, CNRS, 34095 Montpellier, France}
\author{T. Guillet}
\affiliation{Laboratoire Charles Coulomb, Université de Montpellier, CNRS, 34095 Montpellier, France}
\author{S. Bouchoule}
\affiliation{Centre de Nanosciences et de Nanotechnologies, CNRS, Universit\'e Paris-Saclay, France}
\author{J.~Z\'u\~niga-P\'erez}
\affiliation{UCA, CRHEA-CNRS, Rue Bernard Gregory, 06560 Valbonne, France}
\affiliation{MajuLab, International Research Laboratory IRL 3654, CNRS, Université Côte d’Azur, Sorbonne Université, National University of Singapore, Nanyang Technological University, Singapore, Singapore}
\author{D. D. Solnyshkov}
\affiliation{Universit\'e Clermont Auvergne, Clermont Auvergne INP, CNRS, Institut Pascal, F-63000 Clermont-Ferrand, France}
\affiliation{Institut Universitaire de France (IUF), 75231 Paris, France}
\author{G. Malpuech}
\affiliation{Universit\'e Clermont Auvergne, Clermont Auvergne INP, CNRS, Institut Pascal, F-63000 Clermont-Ferrand, France}

\begin{abstract}
Bound states in the continuum (BIC) are of special interest in photonics due to their theoretically infinite radiative lifetime. 
Here, we design a structure composed of a GaN layer with guided exciton-polaritons and a TiO$_2$ 1D photonic crystal slab. The photonic BIC hosted by the photonic crystal slab couples with the excitons of GaN to form a polaritonic BIC with a negative mass. This allows condensation to be reached with a low threshold in a structure suitable for electrical injection, paving the way for room-temperature polariton microdevices. We study in detail how the repulsive interaction between exciton-polaritons affects the condensate distribution in reciprocal space and, consequently, the condensate's overlap with the BIC resonance and, therefore, the condensate lifetime. We study an intrinsic contribution related to the formation of a bright soliton and the extrinsic contribution related to the interaction with an excitonic reservoir induced by spatially focused non-resonant pumping. We then study the peculiar dynamics of the condensation process in a BIC state for interacting particles using Boltzmann equations and hybrid Boltzmann-Gross Pitaevskii equations. We find optimal conditions allowing one to benefit from the long lifetime of the BIC for polariton condensation in a real structure.
\end{abstract}

\maketitle

\section{Introduction}
A large part of Physics is built upon the conservation of quantities, such as energy or momentum, related to symmetries~\cite{Noether1918}. This is the case of the Hamiltonian description of classical mechanics and then of quantum mechanics. However, the irreversible decay of a particle or a mode because of its coupling to a continuum is ubiquitous and, therefore, essential to describe various phenomena. Techniques have been developed, for example in quantum optics, which separate the reversible Hamiltonian dynamics and the dissipative dynamics related to the irreversible decay towards a continuum~\cite{lindblad1976generators}. A somewhat simpler description of multimode non-conservative systems came with the development of non-Hermitian theories, where modes have complex energies with imaginary parts describing their decay~\cite{krol2022annihilation}. The coupling between two such modes hybridizes them but also redistributes~\cite{Bender1998} their imaginary parts $\Gamma_{1,2}$. In the case of a Hermitian coupling, the maximal redistribution occurs at exceptional points (EPs), where the mode decay becomes the mean value of that of the uncoupled modes $(\Gamma_{1}+\Gamma_{2})/2$. It can be zero only in systems with equal gain and loss (PT-symmetric) for the original modes. The study of non-Hermitian systems and in particular of EPs became extremely popular in the last decade \cite{bergholtz2021exceptional}, partly because of the possibility offered to control the decay of the modes. This is essential for many applications including lasing~\cite{peng2016chiral}, and can even lead to fundamentally new concepts, such as the non-Hermitian skin effect~\cite{yao2018edge,kokhanchik2023non}. 

Extra opportunities in this quest of controlling lifetimes came from the so-called dissipative coupling introduced in the 80s~\cite{friedrich1985interfering}. The idea is that two modes are coupled to the same continuum and because of this coupling a transfer of particles from one mode to another can take place with a rate computed as $i\sqrt{\Gamma_1\Gamma_2}$ ~\cite{devdariani1976crossing}. This non-Hermitian term offers a new resonance condition, where one eigenmode of the coupled system becomes entirely non-decaying. This type of mode is called bound state in the continuum (BIC) ~\cite{hsu2016bound}. It was described long ago in the context of quantum mechanics~\cite{vonNeuman1929}, but BICs in photonic systems appear very appealing and commonly present ~\cite{azzam2021photonic}. BICs in photonic crystal slabs or metasurfaces are modes that lie above the light cone but are not coupled to the far field. One should note that other decay mechanisms, such as disorder-induced elastic scattering towards other modes, are not suppressed for BICs, which means that, in reality, their lifetime is finite. In the 2000s, when photonic crystal slabs (PCS) became widely studied, a very important goal was to improve the lifetime of the modes to realize lasers, but the BIC concept had not yet been established. Nevertheless, experimental and theoretical works found that modes with a long lifetime could be realized near the $\Gamma$ point of photonic crystals dispersion~\cite{matsubara2008gan,solnyshkov2011polariton}. Here, the BICs form when two modes of the same polarization and parity couple through the continuum. 

Recently, BICs have been implemented in strongly coupled exciton-photon systems where exciton-polaritons (polaritons) form~\cite{grudinina2023collective,gianfrate2024reconfigurable}. As expected, non-resonantly pumped polaritons were found to accumulate in the state with the longest lifetime, namely the BIC-state, forming a Bose-Einstein Condensate (BEC) \cite{ardizzone2022polariton,riminucci2022nanostructured}. A principal advantage of a polariton condensate versus a photon laser is that the transparency condition required for lasing is not necessary for polariton lasing~\cite{imamog1996nonequilibrium,kavokin2017microcavities}. The only lasing condition for polaritons (which is similar for photon lasers) is that scattering towards a state overcomes its losses, which are expected to be essentially radiative losses for polaritons and that vanish for BICs. Another key difference between polaritons and bare photons is that polaritons are interacting particles. They interact with themselves and with reservoir excitons created by an external pumping. Depending on the sign of the effective mass these interactions, which are typically repulsive, can either tend to delocalize particles~\cite{Pitaevskii} in real space or localize them, forming bright solitons~\cite{tanese2013polariton,pernet2022gap} or simply reservoir-induced localized states \cite{milicevic2018lasing}. In systems with both positive and negative mass states, the localization near the exciton reservoir induced by the pump on negative mass states tends to favor condensation in these states \cite{tanese2013polariton,jacqmin2014direct,milicevic2018lasing}. An important consequence of the formation of either soliton-like BEC or of BEC bound to a spatially-localized exciton reservoir is that it broadens the condensate distribution in $k$-space. In a system containing a BIC, this $k$-space broadening can considerably reduce the average particle lifetime in the BEC. Finally, BIC-based polariton BEC has been demonstrated in GaAs-based structures at low temperature \cite{ardizzone2022polariton,riminucci2022nanostructured}. The potential use of polariton-based devices requires room temperature operation, which is typically implemented either using large-bandgap semiconductors (ZnO \cite{zamfirescu02,li_excitonic_2013,jamadi2018edge}, GaN \cite{malpuech_room-temperature_2002,christopoulos_room-temperature_2007,souissi_Ridge_2022}), organic materials~\cite{kena2010room,plumhof2014room} or perovskites~\cite{su2017room,su2020observation,peng2022room}. In that perspective, GaN-based devices remain the best potential choice since this material offers the best stability as an inorganic semiconductor, and also the possibility to fabricate electrically-injected devices.

In this work, we first propose a design of a realistic BIC-based polariton laser in a GaN/TiO$_2$ polaritonic crystal slab, using finite element methods to solve Maxwell's equations. We then focus on polariton-specific features, particularly the role played by interactions. In the conservative limit we determine the shape of the polariton BEC with and without the reservoir potential. This allows us to analytically compute the average BEC particle lifetime in both cases. We then use the computed lifetimes to model quantitatively the polariton condensation using Boltzmann equations. We find that the advantage of having a BIC is strongly reduced, if not cancelled, when the interaction with a reservoir reduces too strongly the condensate extension in real space, explaining recent experimental results~\cite{riminucci2022nanostructured}. We find that the formation of a soliton provides anyway an intrinsic limitation to the lifetime enhancement provided by the BIC, although the threshold can still be reduced by more than one order of magnitude. At large pumping density, the formation of a soliton provides saturation of the condensate density growth rate versus pumping, which scales in $P^{1/3}$ instead of being linear. Finally, we model the polariton condensation process, solving numerically hybrid Boltzmann-Gross-Pitaevskii equations \cite{solnyshkov2014hybrid}, which allows us to take into account self-consistently the energy relaxation and the inter-particle interactions.

\section{Design of the structure}\label{sec1}
\begin{figure}[tbp]
    \centering
    \includegraphics[width=0.8\linewidth]{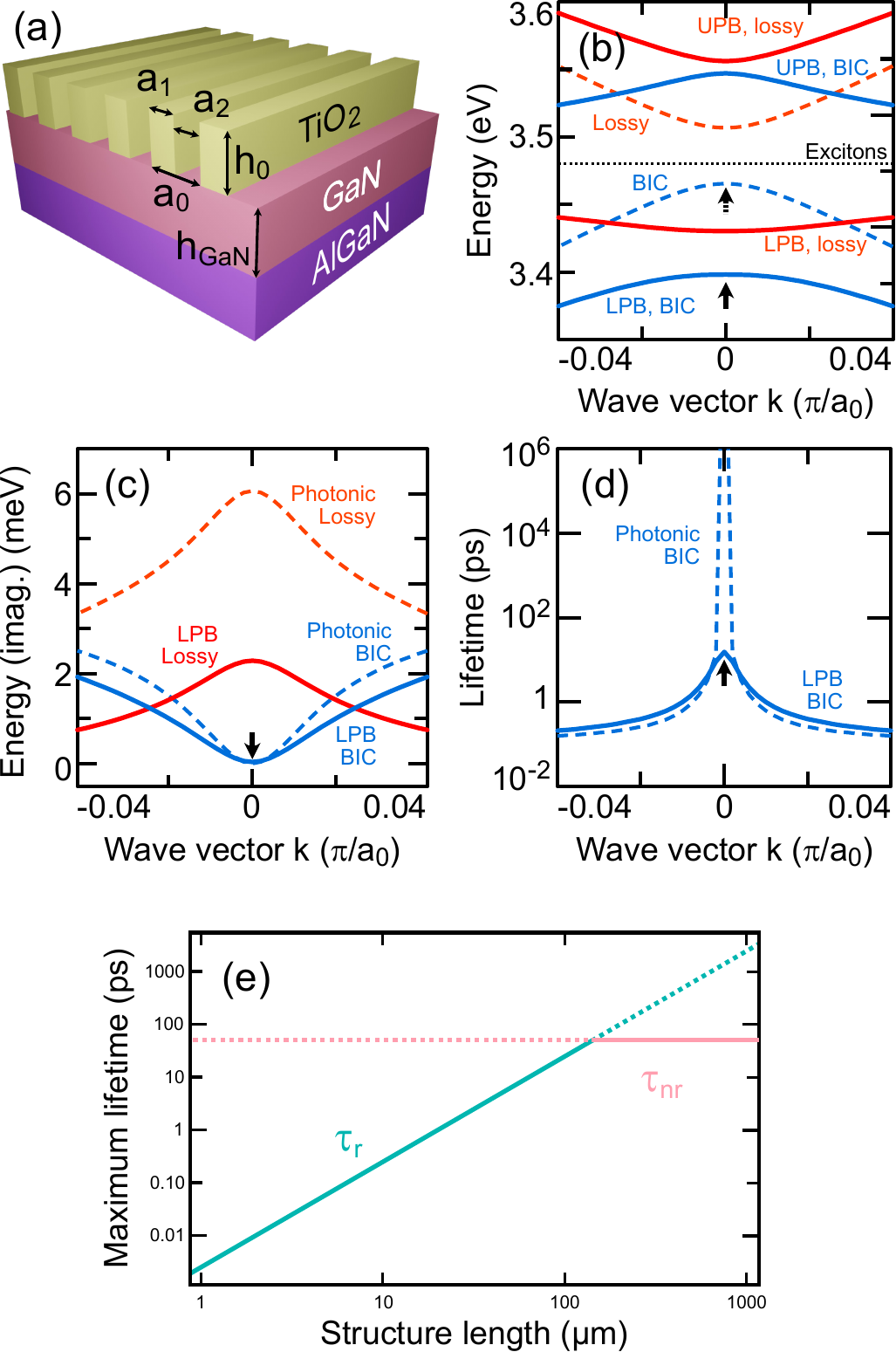}
    \caption{(a) Sketch of the photonic crystal slab structure studied. A two-layer waveguide (TiO$_2$ in yellow and GaN in pink) is separated from the substrate by a cladding layer (AlGaN in purple). The TiO$_2$ layer is etched with a step-like profile (step width $a_1$). (b) Photonic dispersion relation (dashed lines) for the branch containing a BIC (BIC branch, in blue) and the branch containing no BIC (lossy branch, in red). The corresponding polaritonic dispersion relations are plotted in solid lines (red/blue for the polariton dispersion relation calculated from the photonic lossy/BIC branch, respectively). There are two polariton branches (Upper/Lower Polariton Branch) for each photonic branch. The exciton effective energy is indicated as a dotted line. We extract $m$ from the BIC LPB. (c) Imaginary part of the energy of the photonic branches (dashed lines) and lower polaritonic branches (solid lines) versus wave vector. The red/blue curve corresponds to the lossy/BIC branch. We extract $\Lambda$ from the BIC LPB. (d) Photonic/polaritonic (dashed/solid line) lifetime (log scale) against wave vector. (e) Maximum lifetime achievable depending on the structure length $L$. The blue/pink lines indicate the radiative/non-radiative lifetimes. In (b-d), the polaritonic/photonic BIC at $k=0$ is indicated with a solid/dashed black arrow.}
    \label{fig1}
\end{figure}

The structure we study is sketched in Figure~\ref{fig1}(a). It is composed of a TiO$_2$ slab deposited on top of a GaN waveguide. The GaN slab is isolated from the bulk GaN substrate by a cladding layer, that we choose to be Al$_x$Ga$_{1-x}$N, with $x \approx 0.2$, as often done in experiments~\cite{souissi_Ridge_2022}. The TiO$_2$ layer is patterned with a 1D step pattern, forming a photonic crystal slab. Its lattice constant is denoted $a_0$. The width of the steps is $a_1$, while the space between them is $a_2$, with $a_1 + a_2=a_0$.

We simulate the structure in COMSOL and find the dispersion relation of the photonic modes $E_\mathrm{p}(k)$, together with their radiative losses $\Gamma_\mathrm{r}(k)$. The structural parameters we use are $a_0 = 135~$nm, $a_1=80~$nm, $a_2=55~$nm, $h_0 = 50$~nm, and $h_\mathrm{GaN}=100~$nm. The dispersion relation is plotted in Figure~\ref{fig1}(b) (dashed lines). We see that there are two photonic branches, both behaving quadratically for low wave vectors. We call the negative-mass photonic branch the ''BIC'' branch because it contains the BIC (at $k=0$), while the branch at higher energy is called the ''lossy'' branch because the states are lossy at all wave vectors. A dashed arrow indicates the photonic BIC (state) at $k=0$. We can approximate the photonic dispersion relation with a quadratic dispersion of the form:
\begin{equation}
    E_{\mathrm{p}}^\mathrm{BIC/lossy}(k) = E_{0,p}^\mathrm{BIC/lossy} + \frac{\hbar^2 k^2}{2 m_{\mathrm{p}}^\mathrm{BIC/lossy} },
\end{equation}
where $\hbar$ is the reduced Planck's constant and $m_{\mathrm{p}}$ the photonic effective mass. $E_{0,p}$ is the energy of the branches at $k=0$. The lossy branch has a positive mass, while the BIC branch has a negative one in our case. This can be inverted by tuning the parameters, particularly the filling factor $a_1/a_0$. 

In Figure~\ref{fig1}(c), we show the radiative losses (imaginary part of the energy, $\hbar\Gamma$) of the photonic modes. We can see that while the radiative losses of the BIC branch are zero for $k=0$, the losses of the lossy branch are always nonzero. Similarly to the real part of the energy, the imaginary part can be approximated by a square dependence on the wave vector:
\begin{equation}
    \Gamma_{\mathrm{r}}^\mathrm{BIC/lossy}(k) = \Gamma_\mathrm{r0}^\mathrm{BIC/lossy} + \gamma_\mathrm{r}^\mathrm{BIC/lossy} k^2,
\end{equation}
where $\Gamma_\mathrm{r0}$ are the radiative losses at $k=0$. For the BIC branch, this term is zero because the losses vanish at the BIC at $k=0$:
\begin{equation}
    \Gamma_{\mathrm{r}}^\mathrm{BIC}(k) =  \gamma_\mathrm{r} k^2.
\end{equation}

We can calculate the lifetime of the modes from the radiative losses as:
\begin{equation}
    \tau_\mathrm{r} (k) = \frac{1}{2\Gamma_\mathrm{r}(k)},
\end{equation}
and we see that the radiative lifetime of the BIC is theoretically infinite, as can be seen in Figure~\ref{fig1}(d). However, several limitations prevent real BICs from having an infinite lifetime. The first limitation results from the finite size of the structure. Indeed, the size of the full structure $L$ limits the minimal wave vector that can be achieved $k_\mathrm{min}\approx \frac{2 \pi}{L}$.
The radiative lifetime of the BIC branch is infinite only at $k=0$, so since this value is not achievable, the radiative lifetime of an ideal photonic BIC is limited to:
\begin{equation}\label{taur}
    \tau_\mathrm{max}^\mathrm{BIC} = \frac{L^2}{8\pi^2\gamma_\mathrm{r}}
\end{equation}
This is represented in Figure~\ref{fig1}(e) as a green line: the larger the structure, the longer the radiative lifetime, but a real (finite) structure will never exhibit an infinite lifetime. Moreover, non-radiative losses can also play an important role, as we explore in the following.

We designed the structure so that the energies of interest are close to the excitonic resonance of GaN. The photonic modes will thus couple to the excitonic modes and form exciton-polaritons, which are part-light, part-matter quasiparticles. To find the polaritonic dispersion relation out of the photonic dispersion relation containing the lossy branch and the BIC branch, we consider the strong coupling between excitons and photons through the effective strong coupling Hamiltonian:
\begin{equation}\label{matrixSC}
    M_\mathrm{SC}= \left( \begin{array}{*{20}{c}}
E_\mathrm{X}(k) & \rho \hbar \Omega_\mathrm{R}/2 & 0 & 0 \\ \rho \hbar \Omega_\mathrm{R}/2 & E_\mathrm{p}^\mathrm{lossy}(k) & 0 & 0 \\
0 & 0 & E_\mathrm{X}(k) & \rho' \hbar \Omega_\mathrm{R}/2 \\
0 & 0 & \rho' \hbar \Omega_\mathrm{R}/2 & E_\mathrm{p}^\mathrm{BIC}(k)
\end{array} \right),
\end{equation}
where $E_\mathrm{X}(k) \approx E_\mathrm{X} = 3480~$meV at 5~K and 3520~meV at room temperature~\cite{brimont2020strong,mallet2022low} is the effective excitonic resonance of GaN, which groups both A and B exciton contributions. The two different photonic modes have distinct spatial distributions and are coupled to distinct spatial distributions of excitons, which explains the structure of the Hamiltonian.
$E_\mathrm{p}(k)$ are the photonic dispersions found in COMSOL, and we take the Rabi splitting  $\hbar \Omega_\mathrm{R} = 100~$meV for GaN \cite{brimont2020strong}. $\rho$ and $\rho'$ are the fractions of the lossy/BIC modes, respectively, that are located in the GaN layer. Indeed, since the waveguide is composed not only of GaN, but also of a TiO$_2$ photonic crystal slab, which does not contain excitons, we take it into account in the overlap integral between the photon and excitons wave function, which sets the Rabi splitting value, as we did previously~\cite{septembre2023design}. In our design, the value of $\rho$ is typically around 60\,\%, while the value of $\rho'$ is close to 75\,\%. The polaritonic dispersion arising from the strong coupling between excitons and photons is determined by finding the eigenvalues of the matrix $M_\mathrm{SC}$. The polaritonic dispersions read:
\begin{equation}
    E_\pm (k) =\frac{E_\mathrm{X}+E_\mathrm{p}(k)}{2}\pm\frac{1}{2}\sqrt{(\rho\hbar \Omega_\mathrm{R})^2+\left(E_\mathrm{X}-E_\mathrm{p}(k)\right)^2},
\end{equation}
where the $+$ sign corresponds to the upper polariton branches (UPB) with energies above the exciton resonance, and the $-$ sign corresponds to the lower polariton branches (LPB) whose energies lie below the exciton resonance.

We plot the resulting polaritonic dispersion in solid lines in Figure~\ref{fig1}(b). The exciton energy is represented as a dotted black line, which is horizontal because the mass of the excitons is orders of magnitude larger than the effective mass of photons. One obtains four polaritonic branches from the two excitonic and two photonic modes. The UPB and LPB corresponding to the BIC (blue) and lossy (red) branches are shown in the figure. The main branch of interest for us is the BIC LPB, represented as a solid blue line with an energy below the exciton energy. The BIC is located at $k=0$ and indicated by a black arrow. We see that at low wave vectors, the dispersion relation of this branch can be approximated by a quadratic dependence on the wave vector:
\begin{equation}
    E(k) \approx  E_0^\mathrm{BIC} + \frac{\hbar^2 k^2}{2 m },
\end{equation}
where $m=-(1.5\pm 0.1) \times 10^{-6} \, m_0$ is the fitted polariton mass, $m_0$ being the free electron mass. This effective mass is negative. $ E_0^\mathrm{BIC}$ is the energy of the BIC LPB branch at $k=0$.

In the strong coupling regime, the imaginary part of the polariton energy $\Gamma$ can be expressed as~\cite{kavokin2017microcavities}:
\begin{equation}
    \Gamma = C_\mathrm{X}(k) \Gamma_\mathrm{X} + \left (1-C_\mathrm{X}(k) \right ) \Gamma_\mathrm{p}(k),
\end{equation}
where $\Gamma_\mathrm{p,X}$ are the losses of the photonic/excitonic modes respectively, and $C_\mathrm{X}$ is the exciton fraction. This fraction can be modulated by engineering the dispersion relation; more precisely, it depends on the detuning between the exciton energy and the photonic energy at $k=0$. Thus, by engineering the structure (period, etc), one can increase the excitonic part of the polaritonic modes and consequently increase the interactions, which have been experimentally measured in GaN just recently~\cite{gromovyi2023mode}.

The lifetime of the polaritons, as that of the photonic modes, can be fitted by a quadratic dependence on the wave vector for small $k$. The approximate expression is given by:
\begin{equation}
    \hbar\Gamma_{\mathrm{pol}}(k) \approx \hbar\Gamma_0 + \Lambda \frac{\hbar^2 k^2}{2 m},
    \label{poldecrate}
\end{equation}
where $\Gamma_0$ are the constant losses and $\Lambda = -0.27\pm 0.03$ is obtained from a fitting. $\Lambda$ is negative here because $m$ is negative while the losses increase with the wave vector for the BIC branch.

Ultimately, the non-radiative lifetime introduced by the excitons will limit the lifetime of a polaritonic BIC. According to Ref.~\cite{jamadi2016polariton}, the order of magnitude of the non-radiative excitonic lifetime is 50~ps at room temperature in GaN. This means that the longest lifetime possible for a polaritonic BIC in this structure is limited, even in an infinite structure. Figure~\ref{fig1}(e) shows that the lifetime of the BIC is limited by the size of the structure when the structure is small but finally, for large structures, the non-radiative lifetime will limit the polaritonic lifetime. Still, depending on the parameters, very high lifetimes can be observed in realistically large structures. For a structure of  100~$\mu$m length, the lifetime is of the order of 20~ps (corresponding to a quality factor of $Q \approx 20 000$ in a $\lambda/2$ microcavity). This is a very high value at room temperature, even for guided modes~\cite{Ciers2020} (and here, the mode is above the light cone). For larger structures, the lifetime is ultimately limited by the non-radiative lifetime, not by the structure's size.


\section{Influence of the interactions on the lifetime}
In this section, we use the $k$-dependent losses of the BIC to compute the losses of the states hosted by the structure. Indeed, as shown in the previous section, the decay rates (and lifetimes) of modes around the BIC are strongly $k$-dependent. An ideal non-interacting homogeneous condensate would occupy a single state $k=0$ and benefit from the corresponding extremely long lifetime.
However, different mechanisms can lead to a localization of the condensate in real space and thus to a broadening in $k$-space, leading to an increase in the overall condensate decay rate (decrease of lifetime). This contribution can be computed as:
\begin{equation}\label{gammaR}
    \braket{\Gamma_\mathrm{pol,r}}=\frac{1}{N} \int \Gamma_\mathrm{pol,r}(k) |\psi(k)|^2 dk.
\end{equation}
where $\psi(k)$ is the $k$-component of the condensate wave function normalised to the number $N$ of condensed particles and $\hbar\Gamma_\mathrm{pol,r}(k)=\Lambda \hbar^2 k^2/2m$ is the wavevector-dependent part of the polariton decay rate~\eqref{poldecrate}.

There are two localization mechanisms in real space for negative-mass particles with repulsive interactions. The first one is an intrinsic mechanism related to the formation of a bright soliton, whose width is related to $N$. The second mechanism takes place if the pump is non-uniform (localized) in real space. The generated excitonic reservoir repulsively interacts with the condensate, playing the role of an attractive potential. Generally, both mechanisms are present ~\cite{tanese2013polariton}. We first treat this problem analytically, using the conservative Gross-Pitaevskii (GP) equation:
\begin{equation}
    i\hbar\frac{\partial \psi}{\partial t}=-\frac{\hbar^2}{2m}\frac{\partial^2 \psi}{\partial x^2}+     \alpha |\psi|^2\psi +U(x)\psi
\end{equation}
where $\alpha$ is the polariton-polariton interaction constant and $U(x)$ an external potential induced by the exciton reservoir.

So far, the losses have been neglected. We start with the intrinsic mechanism of bright soliton formation (without any external potential $U(x)=0$). The bright soliton solution of the GP equation without external potential reads:
\begin{equation}\label{wfsoliton}
    \psi_\mathrm{s} = \frac{\psi_\mathrm{s}^{(0)}}{\cosh(x/\sqrt{2}\xi)},
\end{equation}
where $\xi$ is the healing length defined as:
\begin{equation}
    \xi = \frac{\hbar}{\sqrt{2 |m \alpha |}\psi_\mathrm{s}^{(0)}}
\end{equation}
and $\psi_\mathrm{s}^{(0)}$ is the value of the wave function at $x=0$, defined as:
\begin{equation}
    \psi_\mathrm{s}^{(0)}=\frac{\sqrt{|m \alpha |} N}{2\hbar},
\end{equation}
which ensures the normalization of the wave function to $N$. Finally, the healing length can be rewritten explicitly as a function of $N$ as:
\begin{equation}
    \xi = \frac{ \sqrt{2} \hbar^2}{|m\alpha | N}.
    \label{xiN}
\end{equation}

This shape allows us to compute analytically the decay rate in the presence of a BIC using formula~\eqref{gammaR} as: 
\begin{equation}\label{losses_soliton}
\braket{\hbar\Gamma_\mathrm{r}} = \Lambda \frac{\alpha^2 m}{12\pi^2\hbar^2}N^2
\end{equation}

This non-linear loss rate will saturate the growth of $N$ after the condensation threshold. Qualitatively, the accumulation of particles in the soliton renders it larger in $k$-space, which increases the decay rate, which then limits the growth of $N$.

We now consider the localization by an external potential induced by the finite size of a Gaussian reservoir created by a non-resonant pumping:
\begin{equation}
    U(x)= U_0 e^{-\frac{x^2}{2\sigma^2}},
\end{equation}
where $U_0$ is the height of the potential "trap" formed by the pump and $\sigma$ is its width in real space. The full width at half maximum (FWHM) of the pump is $2\sqrt{2 \ln (2)} \sigma$.
We approximate the potential by a parabolic profile found from the Gaussian profile by a series expansion:
\begin{equation}
    U \approx U_0 \left ( 1-\frac{x^2}{\sigma^2}\right ).
\end{equation}
The solutions to the Schrödinger equation for negative mass particles in this harmonic potential correspond to the solutions for positive mass particles but with a potential $-U$. The solutions are the well-known states of the harmonic oscillator, which are the products of Hermite polynomials and of a Gaussian function. This harmonic oscillator description is valid only for particle energies within the gap between the BIC and lossy branches. This limits the number of modes confined by the potential and, in practice, we will only consider the two first modes $\Tilde{\psi}_{0,1}$ defined as:  
\begin{equation}
    \Tilde{\psi}_0 = \frac{1}{\sqrt{w\sqrt{\pi}}} e^{-\frac{x^2}{2w^2}},~\Tilde{\psi}_1 = \sqrt{2}\frac{x}{w} \Tilde{\psi}_0.
\end{equation}
where $w$ is the width of the Gaussian wave function. The $p$-state is expected to host a condensate when the energy of the $s$-state becomes larger than the gap and is therefore not confined anymore~\cite{tanese2013polariton,milicevic2018lasing,riminucci2023bose,nigro2023theory}.

Because of the presence of both the interactions and the potential, the Gross-Pitaevskii equation is not exactly solvable. We use the variational approach to find the solution analytically, using the width $w$ of the Gaussian wave function as a variational parameter. We start with a trial wave function of a Gaussian shape, given by $\psi_0 = \sqrt{N} \Tilde{\psi}_0$ where $N$ is the number of particles (note that $\int|\psi_0|^2 dx=N$ while $\int |\Tilde{\psi}_0|^2 dx=1$). We then find the width $w$ that minimizes the total energy of the system $E_\mathrm{tot}$, which is the sum of the kinetic, potential, and interaction energy:
\begin{equation}
    E_\mathrm{tot} = E_\mathrm{kin} + E_\mathrm{pot} + E_\mathrm{int}.
\end{equation}
The minimization condition allows us to write an equation for $w$ that we solve:
\begin{equation}\label{dEtot}
    \frac{\partial E_\mathrm{tot}}{\partial w}=0.
\end{equation}

The different energy terms are calculated from the trial wave function as follows~\cite{Pitaevskii}:
\begin{equation}
    E_\mathrm{kin} = \int -\psi_0^*\frac{\hbar^2}{2 m} \frac{\partial^2 \psi_0}{\partial x ^2}  dx,
\end{equation}
\begin{equation}
    E_\mathrm{pot} = \int |\psi_0|^2 U dx,
\end{equation}
\begin{equation}
    E_\mathrm{int} = \int \frac{\alpha }{2} |\psi_0|^4dx.
\end{equation}
In the end, we find the total energy, written here per particle:
\begin{equation}
    E_\mathrm{tot}/N = \frac{\hbar^2}{4mw^2} + \frac{w^2 U_0}{\sigma^2} + \frac{\alpha N}{2\sqrt{2\pi}w}.
\end{equation}
In the absence of interactions ($\alpha=0$), the energy is minimized for a width~\cite{landau2013quantum,solnyshkov2023analog}:
\begin{equation}
    w_{|\alpha=0} \equiv w_0 = \sqrt{\frac{\hbar \sigma}{\sqrt{|m| U_0}}}.
\end{equation}
The trial function is an exact solution in this case.
The decay rate can be found analytically. It reads:
\begin{equation}
    \braket{\Gamma_\mathrm{r,0}} = \frac{\pi\Lambda\hbar}{2m w^2}.
\end{equation}
A similar analysis can be made for state 1, and the decay rate reads:
\begin{equation}\label{lossespstate}
    \braket{\Gamma_\mathrm{r,1}} = 3 \braket{\Gamma_\mathrm{r,0}}, 
\end{equation}
and more generally, the losses of the state $\mathcal{M}$ is given by:
\begin{equation}
    \braket{\Gamma_\mathrm{r,\mathcal{M}}} = \left ( \mathcal{M}+\frac{1}{2} \right ) \frac{\pi \Lambda \hbar}{mw^2}.
\end{equation}

\begin{figure}[tbp]
    \centering
    \includegraphics[width=0.99\linewidth]{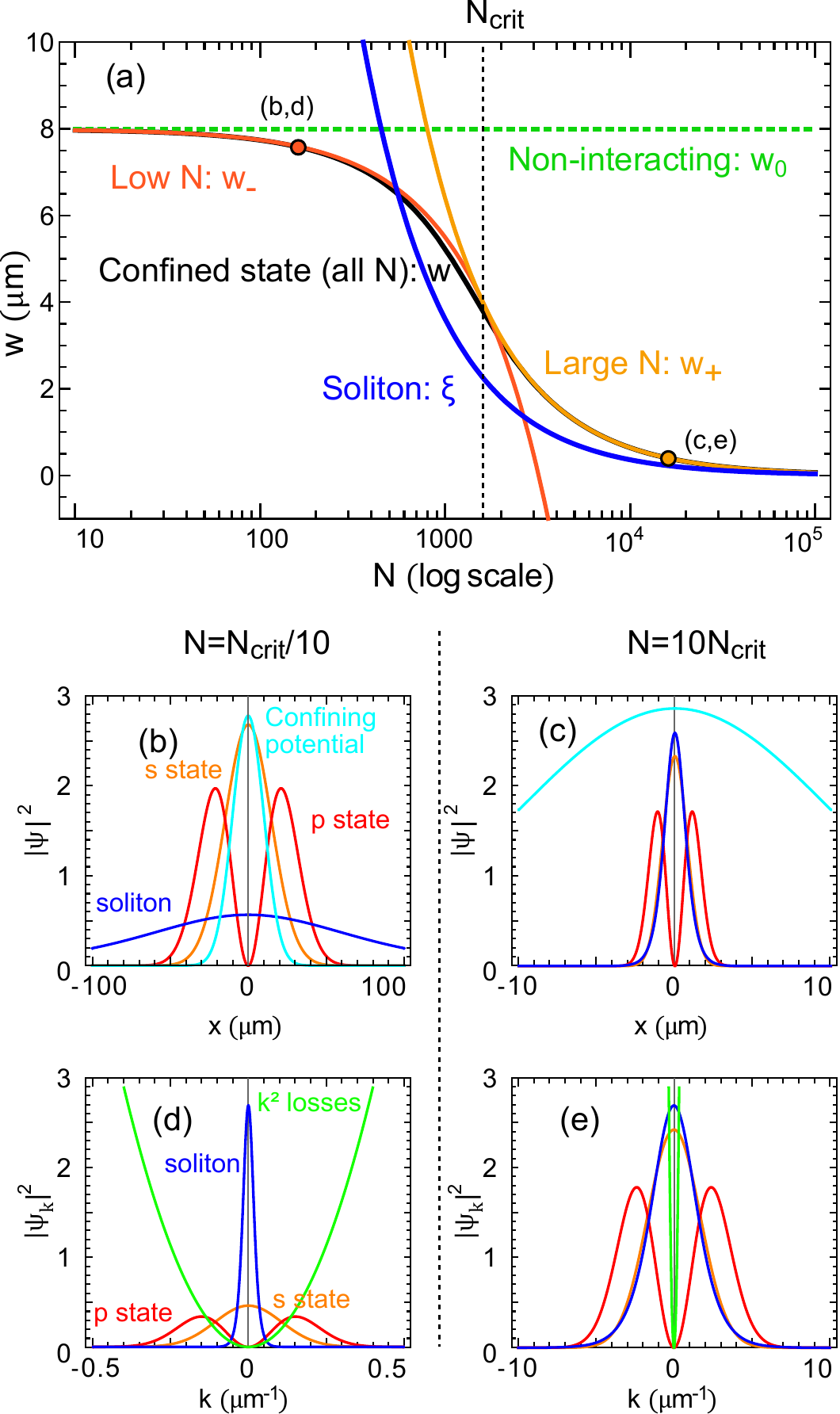}
    \caption{(a) Width of the Gaussian wave function $w$ (black thick line) and healing length of the soliton $\xi$ (blue line) as a function of the number of particles $N$. $w_\pm$ (respectively gold/orange lines) are the approximations of $w$ for large/small $N$, and $w_0$ is the width in the non-interacting case. (b-c) Profile of the wave function in real space for $N=N_\mathrm{crit}/10$ (b) and $N=10N_\mathrm{crit}$ (c). The confining potential (light blue) has a predominant effect at small $N$ but a small effect at large $N$. (d-e) Profile of the wave function in reciprocal space for $N=N_\mathrm{crit}/10$ (d) and $N=10N_\mathrm{crit}$ (e). The $k^2$ losses close to the BIC (green line) give very low losses at low $N$ but high losses at large $N$. In (b-d), the dark blue curve is the soliton wave function and the orange/red curves are the wave functions of the states 0 and 1, respectively.}
    \label{fig4}
\end{figure}

When interactions are present, Eq.~\eqref{dEtot} becomes:
\begin{equation}\label{eqw}
    \frac{U_0}{\sigma^2} w^4 - \frac{\alpha N}{\sqrt{2\pi}} w - \frac{\hbar ^2 }{2m}= 0.
\end{equation}
This is a 4th-order polynomial equation. It is perfectly solvable; however, the expressions of the solutions are quite cumbersome. Among the four solutions only one is real positive, which is the only one we keep since $w$ is a length. To understand the behaviour of the system, we consider analytically the two limits $N\to 0 $ and $N \to \infty$, which give two different expressions for $w$ by expansion in Laurent series. We find that:
\begin{equation}
    w_{|N\to 0} \equiv w_-(N) = w_0 - \frac{\alpha \sigma \sqrt{|m|}}{4\hbar \sqrt{2\pi U_0}}N,
\end{equation}
and
\begin{equation}
    w_{|N \to \infty} \equiv w_+(N) = \frac{\sqrt{2\pi}\hbar^2}{|m\alpha|N}.
    \label{wiN}
\end{equation}
The two limits coincide for a value of $N$ that we denote:
\begin{equation}
    N_\mathrm{crit}=\frac{2\hbar}{\alpha}\sqrt{\frac{2\pi\hbar}{|m|\sigma}\sqrt{\frac{U_0}{|m|}}},
\end{equation}
obtained by solving the equation $w_- = w_+$. This is the critical value for which interactions within the condensate become the dominant mechanism controlling its width. We can therefore write an approximate expression for the width $w(N)$ using the two asymptotic expressions:
\begin{equation}\label{wpm}
    w \approx w_- \mathcal{U}(-(N-N_{\mathrm{crit}})) + w_+ \mathcal{U}(N-N_{\mathrm{crit}}),
\end{equation}
where $\mathcal{U}$ denotes the Heaviside step function.
Figure.~\ref{fig4}(a) shows the width of the 0 mode versus $N$, which is here considered as an external parameter. At low $N$, the mode width is $w_0$ governed by the reservoir potential. The width then decreases and becomes dominated by interaction within the condensate for $N>N_{\mathrm{crit}}$. At very large $N$, the width becomes the same as the one of the bare bright soliton $\xi$. The full solution of \eqref{eqw} is shown as a thick black line, while the two asymptotic solutions combined by \eqref{wpm} are shown by gold and orange lines.

We now consider the second confined state $\psi_{1}$, for which we perform the same type of analysis. We use the spectroscopic notation for the states, as it is often done since the polaritonic trap can be considered an artificial atom. The states 0 and 1 therefore correspond to $s$ and $p$, respectively. 
We remind that state 1 can host the condensate when the energy of state 0 overcomes the band gap between the BIC and lossy branches.
We plot in Figure.~\ref{fig4}(b-e) the real and $k$-space profiles of the different modes for low $N$ (b,d) and large $N$ (c,e). For a low number of particles, the confined states are mostly determined by the width of the confining potential, whereas a free soliton is much broader in real space. On the contrary, for large $N$, the interactions make all modes narrower so that the confining potential (which does not change between (b) and (c)) has only a minor influence on the width of the confined states, which is close to that of the free soliton.

In $k$-space, the trend is inverse: the narrow states in real space are broad in $k$-space. The broadening in $k$-space for all states increases linearly with the growth of $N$ (Eqs.~\eqref{xiN},\eqref{wiN}). Another important aspect is that the $p$-state shows a node at $k=0$ and shows, in all cases, a smaller overlap with the BIC resonance compared with the soliton and the $s$-states. On the other hand, for large $N$, all states are broad in reciprocal space. It means that most of their probability density is located outside the BIC point and, therefore, these states almost do not benefit from the extremely long BIC lifetime.

\begin{figure}[tbp]
    \centering
    \includegraphics[width=0.99\linewidth]{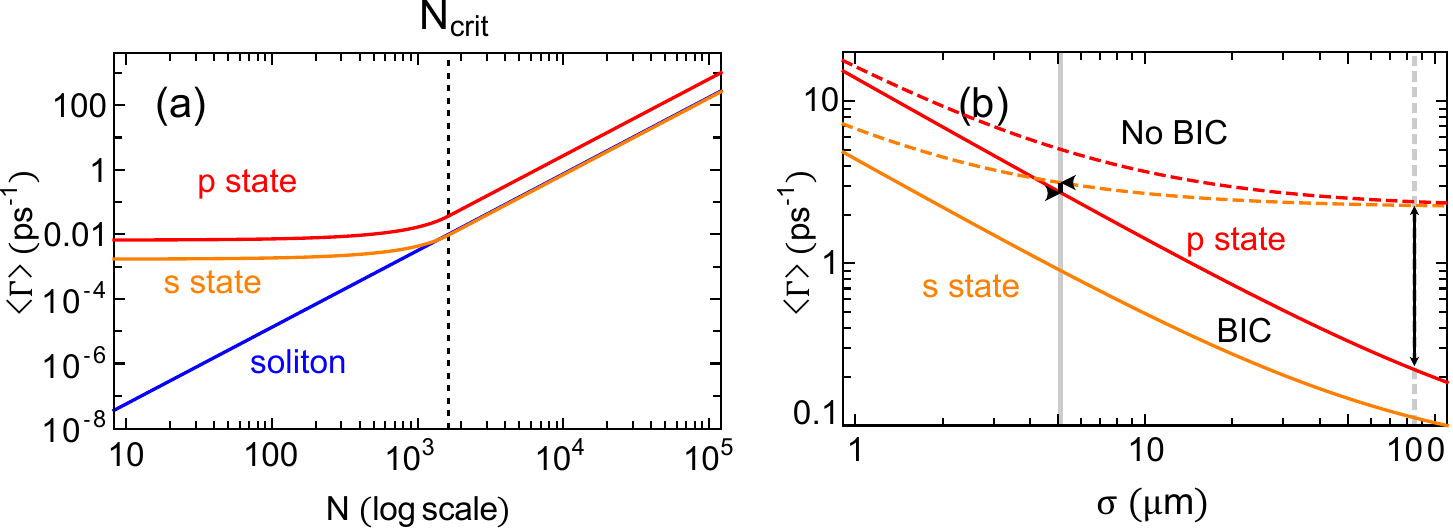}
    \caption{(a) Losses of the soliton (blue line) and $s$- and $p$-states with respect to the number of particles $N$ (in log-log scale). (b) Losses of the $s$-state (orange lines) and $p$-state (red lines) for a band with a BIC (solid lines) and without a BIC (dashed lines). The left (right) grey line corresponds to the situations of Ref.~\cite{riminucci2023bose} (Ref.~\cite{ardizzone2022polariton}), and the black arrows represent the difference between losses of the $s$ state for a non-BIC state and the $p$ state for a BIC state.}
    \label{fig5}
\end{figure}

These features are summarized in Figure.~\ref{fig5}(a), which shows the radiative losses of the soliton, the $s$-, and the $p$-states versus $N$. We observe the $N^2$ dependence of $\Gamma$ for the soliton case. We observe at $N=N_\mathrm{crit}$ the transition for $s$- and $p$-states from a regime where $\Gamma$ is fixed by the pump size to a soliton-like regime dominated by the interactions within the condensate. We observe the constant ratio (equal to 3, see Eq.~\eqref{lossespstate}) for the decay rates between the $s$- and $p$-states. The transition occurs at $N=N_\mathrm{crit}$. All decay rates $\Gamma$ depend on the size of the confining potential $\sigma$. This is shown in Fig~\ref{fig5}b, which also includes a comparison with the band that does not contain a BIC (lossy branch) but instead is characterized by a non-radiative decay rate $\braket{\Gamma_\mathrm{other}}$ (for instance, due to the excitonic resonance when considering exciton-polaritons) and a radiative decay rate $\braket{\Gamma_r}$ which does not depend on $k$ (for instance, the losses of a usual mode above the light cone), so that:
\begin{equation}
    \braket{\Gamma} = \braket{\Gamma_\mathrm{r}} + \braket{\Gamma_\mathrm{other}}
\end{equation}


We observe that for a small reservoir size, the condensate is considerably wider in $k$-space with respect to the range where BIC modifies the decay rate. This situation corresponds approximately to the observation of Ref.~\cite{riminucci2023bose}, where the pump size is approximately $\sigma \approx 5\,$µm. On the contrary, when the pump size increases, the condensate wave function expands in real space and localizes in $k$-space, overlapping better and better with the BIC resonance. In that case, a band with a BIC shows a much smaller decay rate. This is the regime that was demonstrated in Ref.~\cite{ardizzone2022polariton}, where the pumping spot has a width $\sigma\approx 80\,$µm. To summarize, using a small pumping spot prevents benefiting from the high lifetime of a BIC.

\section{Influence of the interactions on the condensation threshold}
In the previous section, we found analytical formulas for the losses of the soliton and $s$- and $p$-states. We show that the losses depend on the number of particles $N$.
In this section, we use the previously found density-dependent losses to solve the semi-classical Boltzmann equations \cite{kavokin2003cavity}, which describe the evolution of the population of the condensate and reservoir states with time. 
In the previous sections, the system was assumed to be 1D and uniform in the transverse direction. Here, we need to take into account a normalization surface in order to operate with the populations and scattering rates. Experimentally, operating with a finite-size pump simply leads to additional non-radiative losses due to the repulsion of the particles from the excitation spot due to the interactions. 
The simplified 2-state model used in this section does not allow us to take into account the details of the relaxation processes that depend on the dispersion relation. Its goal is to focus on the effects of the density-dependent lifetime in the condensate, which is due to the soliton formation.
The two coupled equations read:
\begin{equation}
    \dot{N}_\mathrm{c} = W_\mathrm{LO} N_\mathrm{x}\left(1+N_\mathrm{c} \right )  - N_\mathrm{c}\Gamma_\mathrm{c}(N_\mathrm{c}),
\end{equation}
\begin{equation}
    \dot{N}_\mathrm{x} = -W_\mathrm{LO} N_\mathrm{x}\left(1+N_\mathrm{c} \right )  - N_\mathrm{x}\Gamma_\mathrm{x} + P,
    \label{secondBoltzmann}
\end{equation}
where $N_\mathrm{c,x}$ are, respectively, the numbers of particles in the condensate/reservoir, $\Gamma_\mathrm{c,x}$ are the condensate/reservoir losses, and $P$ denotes the pumping strength. $W_\mathrm{LO}$ describes the relaxation of excitons through LO phonons, which we assume to be the dominant relaxation mechanism. 
Indeed, the design of the structure was optimized in order to have a resonant enhancement for the LO-phonon assisted relaxation: $E_0^\mathrm{BIC} = E_\mathrm{X} + \frac{3}{2} k_\mathrm{B} T - E_\mathrm{LO}$ at $T=300 K$. $W_\mathrm{LO}$ is estimated from Ref.~\cite{kavokin2003cavity} to $W_\mathrm{LO}=6\times 10^9$s$^-1$ for the surface we consider. We neglect the dependence of the scattering rate on the spatial shape of the condensate.  We solve the Boltzmann equations over time until a stationary configuration is obtained. These simulations are repeated for different values of pumping $P$. This allows us to obtain the stationary population of the condensate for each pumping.

\begin{figure}[tbp]
    \centering
    \includegraphics[width=0.99\linewidth]{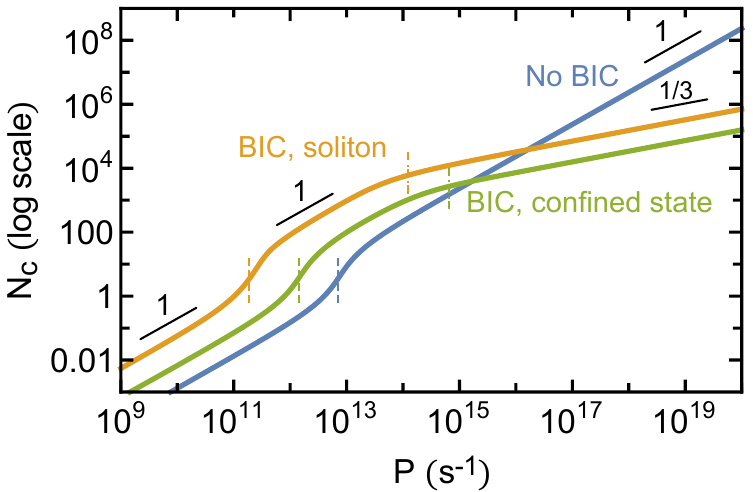}
    \caption{Number of particles in the condensate $N_\mathrm{c}$ versus pumping the strength $P$. The blue line corresponds to the situation without a BIC, the gold line to the situation with a BIC and a free soliton (no confinement), and the green line the situation with a BIC and confined states (the $p$-state in this case). Short vertical dashed lines indicate the condensation thresholds in each situation. Short vertical dashed-dotted lines indicate the slope breaks (1 to 1/3) due to the presence of the BIC.}
    \label{fig6}
\end{figure}

We consider three different cases, which we plot in Figure~\ref{fig6}. In the first case, we consider a configuration where the condensate is not a BIC. The losses for this state (both radiative and non-radiative) are assumed to be $k$-indepedent $\Gamma_\mathrm{c} = \braket{\Gamma_\mathrm{nr}} + \braket{\Gamma_\mathrm{r,0}}$. When increasing the pumping strength, we see that the blue curve exhibits a threshold around $P\approx 10^{13}$~s$^{-1}$, but the slopes before and after the threshold are the same, $N_c\sim P$. 
For large $N_\mathrm{c}$, the stationary solution can be found analytically as:
\begin{equation}
    N_\mathrm{c} \approx \frac{\hbar P}{\Gamma_\mathrm{c}} - \frac{\Gamma_\mathrm{x}}{W_\mathrm{LO}}.
\end{equation}

The two other curves show the solutions when the system hosts a BIC. The golden curve represents the solution without confining potential, where the condensate wavefunction is the one of a soliton. We see that using a BIC instead of an ordinary lossy state reduces the threshold required for condensation by approximately two orders of magnitude (the threshold is estimated to be reached for $N_\mathrm{c} \approx 3$). Before and just after the threshold, the condensate population evolves linearly versus $P$, as in the case without the BIC. However, far above threshold, when $N_\mathrm{c}\approx 10^4$, the dependence changes due to the formation of the soliton, whose losses scale as $N_\mathrm{c}^2$, as shown in Eq.~\eqref{losses_soliton}.
In that regime, keeping only leading terms in $N_\mathrm{c}$ one can find the asymptotic dependence of $N_\mathrm{c}$ versus $P$ as
\begin{equation}
    N_\mathrm{c} \approx \left(\frac{\hbar P}{\beta}\right)^{1/3},
\end{equation}
which corresponds to the 1/3 slope we observe in the numerical results at large $N_\mathrm{c}$. 

Finally, the green curve is obtained when considering a BIC with both confinement by an exciton reservoir and self-interactions. In this case, the losses are given by Equation~\eqref{lossespstate}, and the width is taken from the solution of Equation~\eqref{eqw} (the analytical exact solution, not the approximate solutions). We take the losses of state 1 (the $p$-state) in order to reproduce the situation studied experimentally in Ref.~\cite{riminucci2023bose}. The solution for state 0 ($s$-state) is not plotted because it is very similar to the one we found considering only the soliton without confinement. Indeed, for large $N_\mathrm{c}$, the losses of state 0 are close to the soliton losses, and for small $N_\mathrm{c}$, the solutions of Boltzmann equations are less influenced by $N$-dependent losses. We see that the green curve is an intermediate case between the soliton located at a BIC and a usual configuration with a BIC: the threshold is higher than with the BIC and without confinement, but the interactions limit the growth of $N_\mathrm{c}$ at large values with the same power law. Modifying the width of the confinement potential allows changing the threshold value: the narrower the potential, the higher the threshold.

To conclude, in this section we have shown that the BIC can be beneficial to reduce the threshold. However, the interactions deeply modify the behavior at large pumping strengths/number of particles because the dependence between the two parameters becomes sub-linear. Finally, the size of the reservoir can reduce the effects of the BIC.

\section{Numerical simulations of polariton condensation with Gross-Pitaevskii equation}
In this section, we solve numerically the Gross-Pitaevskii equation to verify qualitatively the validity of the hypotheses used and results obtained in the previous sections. 
The studies performed in the previous sections were based on the hypothesis that the shape of the condensate wave function is a solution to the conservative GP equation. The role of the pumping and lifetime is to set the number of particles in the condensate, which is a wave function parameter. We did not consider that the $k$-dependent lifetime itself could deeply affect the shape of the condensate wave function. In order to take into account on an equal footing interaction and lifetime effects, we solve numerically the hybrid Boltzmann Gross-Pitaevskii equation, including the scattering from a localized excitonic reservoir and $k$-dependent lifetime, similar to what is usually done to describe energy relaxation~\cite{Wertz2012prop,solnyshkov2014hybrid}. The goal here is to check the validity of the analytical predictions for the shape of the condensate and simulate the emission pattern observed in experiments, and not to obtain a quantitative estimate of the threshold, which is why we eliminate the equation for the reservoir.
The equation reads:
\begin{widetext}
\begin{equation}\label{GPBECBIC}
    i \hbar \frac{\partial}{\partial t} \psi(x,t) = \left ( (i \Lambda - 1) \frac{\hbar^2}{2 m}\frac{\partial^2}{\partial x^2}   + \alpha \left(| \psi (x,t) |^2 +2 n_\mathrm{x}(x)\right) + i\hbar( W_{\mathrm{LO}}n_{\mathrm{x}}(x)e^{-N_{\mathrm{x}}/N_\mathrm{sat}}-  \Gamma_0)  \right ) \psi (x,t)+\chi(x,t).
\end{equation}
\end{widetext}
where $m = -(1.5\pm 0.1)\, 10^{-6} m_0$, $\Lambda = - (0.27\pm 0.03)$ as obtained in the first section. $n_\mathrm{x}(x)$ is the reservoir density, which we assume to keep the spatial profile of the pumping laser and to be constant in time. The reservoir depletion is taken into account via the saturated gain term.
We add a noise term $\chi(x,t)$ to simulate the spontaneous scattering into the polaritonic modes from the reservoir. 
The filling of the condensate from an excitonic reservoir generated by a non-resonant optical pumping is assumed to be assisted by the LO phonons through the term containing the LO phonon scattering matrix element $W_\mathrm{LO}$. The saturation is controlled by the total number of particles in the reservoir $N_\mathrm{x}=\int n(x)\,dx$ and its saturation value $N_\mathrm{sat}$, which in the Boltzmann equations is controlled by the pumping $P$ via the second equation~\eqref{secondBoltzmann} that we eliminate adiabatically here. The interaction constant $\alpha$ describes polariton-polariton interaction. A twice larger value for the reservoir excitons is taken in the hypothesis of zero detuning operation. Here, the normalization area for both $\alpha$ and $W_\mathrm{LO}$ is taken equal to the spatial grid size $2~\mu$m times the transverse pump size $20~\mu$m. The main difference between our design and the one presented in Ref.~\cite{wu2023room} is that our BIC is a polaritonic BIC and not just a photonic BIC, which enlarges the potential applications of our proposal compared to this previous realization. The goal of these simulations is to demonstrate that the interactions lead to the formation of a soliton, that its reciprocal-space image corresponds to the one observed experimentally, and that its shape is very close to the one considered analytically.

\begin{figure}[tbp]
    \centering
    \includegraphics[width=0.99\linewidth]{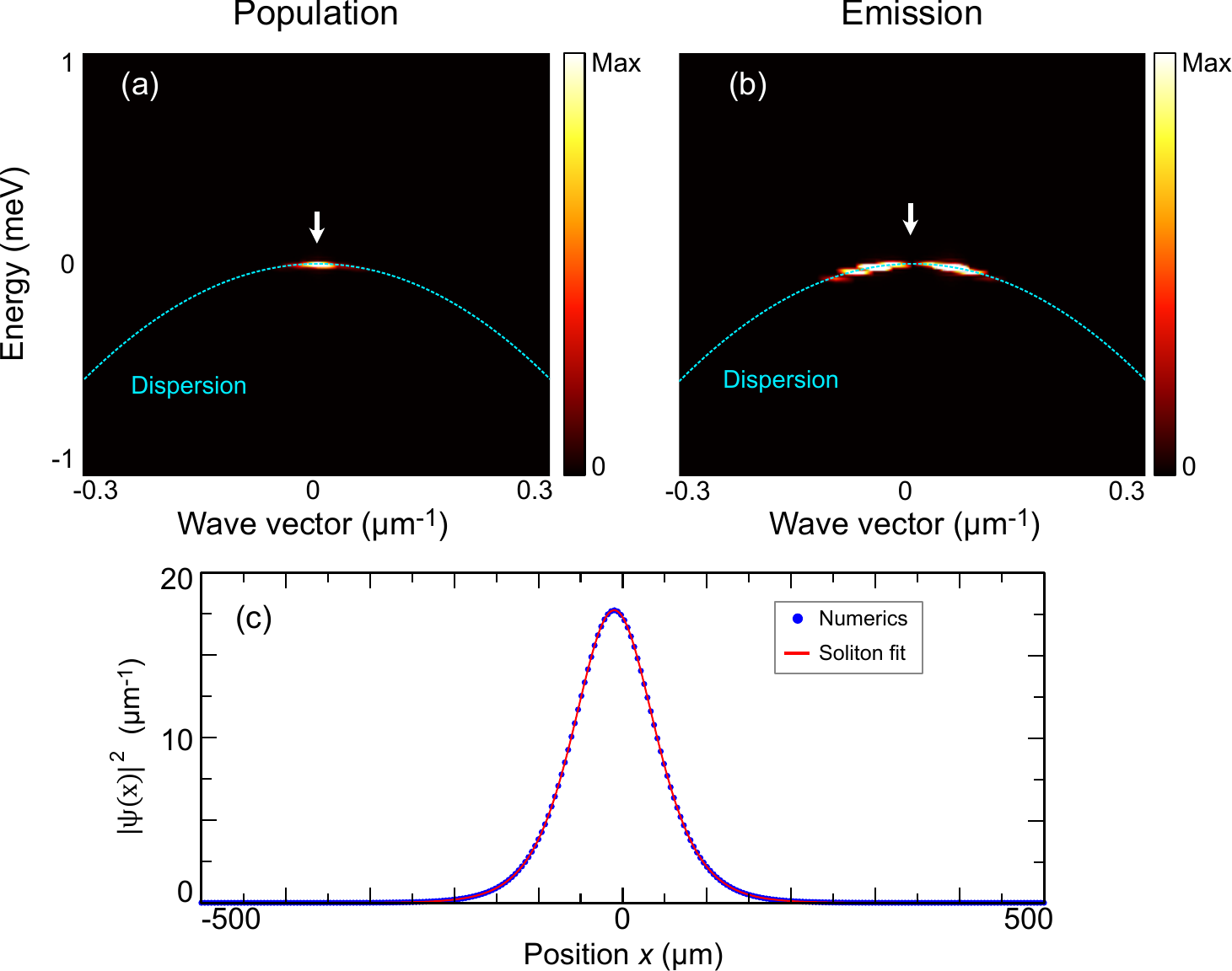}
    \caption{Polariton Bose-Einstein condensate in a BIC. (a,b) Numerical results of the simulation of the Gross-Pitaevskii equation~\eqref{GPBECBIC} for the BIC LPB (containing a BIC at $k=0$). (a) Population in the reciprocal space. Note the unique intensity peak at the maximum of the dispersion (at the BIC). (b) Numerically calculated emission in the reciprocal space. Note the two peaks surrounding $k=0$. In both panels, the dispersion is indicated as a cyan dashed line, and a white arrow shows the position of the BIC (state) in reciprocal space. (c) Numerically obtained distribution of particles in real space (blue points) after the stationary regime is reached and fitted by a bright soliton wave function (red line). Note the excellent agreement.}
    \label{fig3}
\end{figure}

We show the results of numerical simulations based on Eq.~\eqref{GPBECBIC} in Figure.~\ref{fig3} using a homogeneous reservoir profile. The time-dependent simulation is run until a stationary solution is achieved. Above threshold (which in our simulations occurs well below the Mott density), a condensate forms in the ground state of the system, but the interactions lead to its localization in real space and spreading in reciprocal space, as expected. Panel (a) shows the condensate distribution in reciprocal space $|\psi(k)|^2$. We can see that the population is maximal at the top of the dispersion (the dispersion is indicated by a cyan dashed line), which is precisely the position of the BIC. Indeed, the BIC has fewer losses than any other mode, so condensation occurs there.

In most of the systems exhibiting polariton condensation, a higher population corresponds to a higher intensity of emitted light (but even for usual polaritons, one has to take into account the photonic fraction to determine the polariton population from the emission intensity~\cite{kasprzak2006bose}). In the case of BICs, the population is very different from the emission. Indeed, the light emitted by the structure is proportional to the losses: the light that is collected is the light that escaped the waveguide and impinges onto the detector. The emission $\Gamma_\mathrm{r}(k) |\psi(k)|^2$ is plotted in panel (b). We can see two peaks close to the maximum of the dispersion, corresponding to the experimental observation of the polariton condensation at the BICs~\cite{ardizzone2022polariton}. The population is huge in the BIC, but the emission is very low precisely because the lifetime is very high, larger than modes close in reciprocal space, which still have a very high population. In the end, we only see two peaks of emission close to the BIC, but not the BIC itself. 

Our numerical simulations with the Gross-Pitaevskii equation confirm the formation of bright solitons due to the interactions. We plot in Figure~\ref{fig3}(c) the distribution of particles in real space calculated numerically after condensation and after the system has stabilized (the condensation occurs after 20~ps and the picture is taken after 15~ns). We see that there is one peak (blue numerical points) that is well-fitted by the wave function of a bright soliton (red line), as given by Eq.~\eqref{wfsoliton}. The shape corresponds very well to the profile found in the conservative approximation in the previous sections, which confirms the validity of our analytical considerations. Note that the formation of solitons and their interplay with BICs in a different (resonant) pumping configuration have been theoretically discussed in Refs.~\cite{dolinina2021dissipative,dolinina2021interactions}. 
However, we observe numerically an additional effect. When the number of particles exceeds a critical value, the system becomes unstable and does not converge to a stationary solution. This may lead to the formation of several solitons that exhibit periodic oscillations, without merging into a single soliton. According to our analysis, this is not due to the Kibble-Zurek mechanism~\cite{kibble1976topology,zurek1985cosmological}, which has already been proposed to explain different effects in polaritonic structures~\cite{solnyshkov2016kibble,solnyshkov2021kibble,solnyshkov2022domain}. The effect is rather due to an analogue of modulational instability, but in a driven-dissipative system. A detailed study of this phenomenon exceeds the scope of the present article and requires further investigations.

\section{Conclusions}
To conclude, we design a structure for polariton Bose-Einstein condensation (polariton lasing) based on a 1D BIC in a TiO$_2$-GaN waveguide. We predict a possible realization of room temperature polariton Bose-Einstein condensation in the BIC state. This could be implemented using state-of-the-art experimental techniques. It would allow obtaining an electrically-injected room-temperature polariton laser with a low lasing threshold. We also investigate how to benefit from the high lifetime of the BIC. It requires working with the largest structure possible to access the smallest wave vector and the highest lifetimes, and the pumping spot needs to be sufficiently large as well, to prevent confinement effects from reducing the lifetime of the BIC. Soliton formation due to the interactions can increase the losses above threshold. An interesting perspective is to consider condensation of interacting particles in 2D-BIC.

\begin{acknowledgements}
We thank H. S. Nguyen and X. Letartre for participating in the early development of this work, P. Kokhanchik for contributing to the supervision of I. Fudjo, and E. Cambril and J.-Y. Duboz for useful discussions.
This research was supported by the ANR Labex GaNext (ANR-11-LABX-0014), the ANR program "Investissements d'Avenir" through the IDEX-ISITE initiative 16-IDEX-0001 (CAP 20-25), the ANR project "NEWAVE" (ANR-21-CE24-0019) and the European Union's Horizon 2020 program, through a FET Open research and innovation action under the grant agreement No. 964770 (TopoLight).
\end{acknowledgements}

\bibliography{biblio} 
\end{document}